\documentclass[12pt]{article}
\begin{document}
\title{Signature of Fermi Surface Jumps in Positron Spectroscopy Data}
\author{Gh. Adam$^{a,}$\thanks{Senior Associate.
E-mail: adam@ictp.trieste.it} $^{),\; b,}$%
\thanks{Permanent address. E-mail: adamg@roifa.ifa.ro}$~^)$
and S. Adam$^{b,}$\thanks{E-mail: adams@roifa.ifa.ro}$~^)$\\
\smallskip
$^a$ The Abdus Salam International Centre for Theoretical Physics,\\
P. O. Box 586, 34100 - Trieste, Italy\\
\smallskip
$^b$ Department of Theoretical Physics,\\
Institute of Physics and Nuclear Engineering\\
P. O. Box MG-6, RO-76900 Bucharest-M\u{a}gurele, Romania
       }
\date{}
\maketitle

\begin{abstract}
\noindent
A subtractionless method for solving Fermi surface sheets ({\tt FSS}), from
measured $n$-axis-projected momentum distribution histograms by
two-dimensional angular correlation of the po\-si\-tron-electron
annihilation radiation ({\tt 2D-ACAR}) technique, is discussed.
The window least squares statistical noise smoothing filter described in
Adam {\sl et al.}, NIM A, {\bf 337} (1993) 188, is first refined such that the
window free radial parameters ({\tt WRP}) are optimally adapted to the data.
In an ideal single crystal, the specific jumps induced in the {\tt WRP}
distribution by the existing Fermi surface jumps yield straightforward
information on the resolved {\tt FSS}.
In a real crystal, the smearing of the derived {\tt WRP} optimal values, which
originates from positron annihilations with electrons at crystal imperfections,
is ruled out by median smoothing of the obtained distribution, over symmetry
defined stars of bins.
The analysis of a gigacount {\tt 2D-ACAR} spectrum, measured on the archetypal
high-$T_c$ compound $YBa\sb{2}Cu\sb{3}O\sb{7-\delta}$ at room temperature,
illustrates the method. Both electronic {\tt FSS}, the ridge along
$\Gamma X$ direction and the pillbox centered at the $S$ point of the first
Brillouin zone, are resolved.
\end{abstract}

\section{Introduction}
\label{sec:intro}
While the technique of two-dimensional angular correlation of the
positron-electron annihilation radiation ({\tt 2D-ACAR})
\cite{berko83,berko89,peter89},
was successfully used to resolve the Fermi surface topology in various classes
of materials (see, e.g.,
\cite{chan92}
for a recent review), the attempts to use it in high-$T_c$ superconductors met
huge difficulties stemming from various sources: the occurrence of a fraction
of Fermi electrons which is significantly lower than in normal metals, the
layered crystalline structures of the samples, the high proportion of
crystal imperfections in the single crystals under investigation. 

A straightforward consequence of the occurrence of a layered structure, which
prevents the uniform distribution of the thermalized positrons inside the
sample, is the limitation of the usefulness of the {\tt 2D-ACAR} technique to
the resolution of the electronic Fermi sheets only. The occurrence of a small
fraction of Fermi electrons results in very weak Fermi surface jumps in the
measured momentum density, which can thus be easily missed unless the crystal
quality is very high and the accumulated statistics is large enough
to get the small Fermi surface jumps resolved.

The resolution of the Fermi surface in high-$T_c$ superconductors is a
challenge irrespective of the used technique, however. Each of the techniques
used so far ({\tt 2D-ACAR}, de Haas-van Alphen, or angle-resolved
photoemission spectroscopy (see, e.g.,
\cite{chan92}
and references quoted therein), has been able to resolve only part of the
existing Fermi surface sheets, such that the final representation of the
Fermi surface came from the superposition of essentially disjoint pieces of
information
\cite{pankal94}.

In this paper, we consider the identification of the Fermi surface jumps from
a {\tt 2D-ACAR} histogram which records the projection, along a principal
crystallographic axis ${\bf n}$, of the momentum density coming from
zero-spin positron-electron pairs. Such a histogram will be called thereafter
an {\sl $n$-axis-projected histogram}.

If the single crystal used in a {\tt 2D-ACAR} experiment would be perfect, then
the statistical noise smoothing by means of an appropriate method (see, e.g., 
\cite{adam93a,hoffm93a,west95}),
possibly combined with a method of subtracting the radially isotropic
component, would resolve the characteristic Fermi surface jumps.

In fact, there is an important discrepancy between the experimentally measured
and theoretically computed momentum distributions. In the range of low momenta,
the latter, which are calculated under the assumption of perfect crystalline
periodicity, are found to be sensibly smaller than the former
\cite{hoffm91}.
The origins of this discrepancy have been traced back to the positron
annihilation with electrons at crystal imperfections
\cite{smed92,pankal94}.
A quantitative estimate
\cite{pankal94}
suggested that more than one third of all annihilations belong to this
category.

To rule out the spurious effects coming from crystal imperfections, either
their contribution is to be explicitly subtracted from the data (as done,
e.g., by the Argonne group
\cite{smed92,pankal94}),
or the smoothing method is to be insensitive to the occurrence of
crystal imperfections.

Here we report a method of the second kind. There are two basic ingredients
of this method: (i) the derivation of {\sl optimally adapted to the data\/}
radial parameters of the constant-weight window least squares ({\tt CW-WLS})
smoothing method reported in
\cite{adam93a}
and (ii) removal of crystal imperfection dependence of the obtained values by
{\sl median smoothing over stars of symmetry relating the bins of the
histogram}.

The occurrence of Fermi surface jumps in the data results in jumps in the
optimal radial parameters of the {\tt CW-WLS} filter. The presence of crystal
imperfections blurs the occurring jumps. The median smoothing restores the
jumps, if any.

Essential for the success of such an analysis is the operation of
{\sl histogram redefinition from the laboratory frame\/} ({\tt LF}), 
$Op_x^Dp_y^Dp_z^D$, (which is defined by the setup and within which the data
acquisition is performed) {\sl to the crystal frame\/} ({\tt CF}) (which is
identified with $\Gamma p_xp_yp_z$, the canonical reference frame of the first
Brillouin zone of the crystal
\cite{bradl72} 
and within which the various steps of the off-line analysis of the spectrum are
legitimate). Here, this preliminary problem is assumed to be solved along the
lines described in
\cite{adam95}
Within the procedure described in these papers, the assumption (following from
the experiment) of a momentum projection along a principal crystallographic
axis is duly checked, such that the expected symmetrization operations are
justified and hence the inclusion of {\sl stars\/} of bins into the smoothing
procedure of the median method is a valid operation.

The paper is organized as follows. In Section~\ref{sec:cwwls}, the
{\tt CW-WLS} smoothing method is briefly reviewed and the criteria which
serve to the definition of data adapted radial parameters of the smoothing
windows are discussed. The description of the median smoothing is done in
Section~\ref{sec:medsm}. An illustration on experimental data is reported in
Section~\ref{sec:appls}.
\section{{\tt CW-WLS} Statistical Noise Smoothing with\\
Optimal Radial Parameters}
\label{sec:cwwls}
\subsection{Local smoothing windows}
\label{sec:lsw}

Let ${\widetilde H} = ({\widetilde h}_{ij})$ denote the raw {\tt LF}
{\tt 2D-ACAR} histogram of interest, corrected for finite detector aperture
and local variations of the detector sensibility
\cite{adam93a}.
The statistically relevant information within ${\widetilde H}$ {\sl sharply
decreases\/} when one goes from the histogram center towards its
borders.  This feature is preserved in the {\tt CF} histogram $H = (h_{kl})$,
obtained from ${\widetilde H}$ under proper discretization in the
$\Gamma p_xp_yp_z$ frame
\cite{adam95}.

As a consequence, the histogram bins can be divided into two classes:
{\sl central\/} and {\sl border\/} bins. The separation line between the two
bin classes is somewhat arbitrary. For instance, if we decide to restrict the
analysis to the projection of the first Brillouin zone onto the histogram
plane ({\tt 1BZ}), then the central bin area has to include both {\tt 1BZ}
and a surrounding neighborhood of it, half {\tt 1BZ} wide say, to get
detailed information on the momentum distribution around both sides of the
{\tt 1BZ} boundaries.
If, however, the decision is taken to include the maximum possible histogram
area into analysis, an appropriate definition of the central bin manifold
includes the largest possible histogram area $D_\Lambda$ related by a
similarity transform, of factor $(2 \Lambda + 1) \times (2 \Lambda + 1)$ to
the $D_0$ area of {\tt 1BZ}. (Typically, $\Lambda$ equals two or three,
depending on the setup.)

In
\cite{adam93a},
a least squares fit of noise-free data to the noisy $H$ data was proposed to
be performed, at the fractionary coordinates $(\xi, \eta)$ of interest inside
{\sl each\/ $(K, L)$-th central bin} (where the quantities $\xi$ and $\eta$
may vary from bin to bin), by means of {\sl local approximating surfaces},
$C_{KL}$, characterized by the following two features:
\begin{enumerate}
  \item{
     Each $C_{KL}$ is to consist of an integer number of bins. (This
     requirement follows from the {\sl data discretization\/} into square bins,
     that is, into finite regions inside which the structural details of the
     momentum distribution have been averaged out --- and hence lost ---
     within the process of data acquisition. The bin size $a_D$ is taken
     henceforth for the unit length of the distances in the histogram plane.)
       }
  \item{
     In the limit of an in-plane {\sl continuous\/} point distribution, the
     approximating surface $C_{KL}$ is to be {\sl circular\/} around the
     point of coordinates $(K, L)$. (Such a choice shows the {\sl largest
     linear dimensions under the smallest area} and it fits {\sl isotropically}
     the various possible neighbourhoods. It is thus expected to secure, among
     the possible {\tt 2D} shapes, the {\sl least} $L_2\/$ norm
     departure from {\sl arbitrary\/} input surfaces.)
       }
\end{enumerate}
As a consequence, to perform statistical noise smoothing, we draw around
{\sl each\/ $(K, L)$-th central bin} a {\sl smoothing window} $C_{KL}$
of {\sl quasi-circular shape\/} which includes inside it all the bins
the centres $(\kappa, \lambda)$ of which satisfy the inequality
\begin{equation}
  (\kappa - K)^2~+~(\lambda - L)^2~\leq~(2 r + 1)^2/4,
\label{cklr}
\end{equation}
where the quantity $r$ denotes the {\sl window radial parameter}.
\subsection{{\tt CW-WLS} smoothing formula}
\label{sec:smf}

To accommodate both the data discretization into bins and the possibility to
predict noise-free values {\sl inside the bins}, the approximating space of
noise-free data is spanned by a basis set of polynomials of continuous
variables, $P_m (x, y)$, orthogonal over $C_{KL}$. In what follows, it is
assumed that the basis polynomials are of at most 3-rd degree in each of the
variables $x$ and $y$.

Over a local window $C_{KL}$, the standard deviations $\sigma_{K+k,L+l}$
associated to the elements $h_{K+k,L+l}~,~(K+k, L+l)~\in~C_{KL}$ of $H$ show
little variation from $\sigma_{KL}$, the standard deviation of $h_{KL}$
at the center of the window
\cite{adam93a,adam93c}.
We may therefore assume {\sl constant bin weights},
$\sigma_{K+k,L+l} = \sigma_{KL}$, within each smoothing
window~(\ref{cklr}), a hypothesis which results in a {\sl constant weight
{\tt WLS} smoothing formula\/} of radial parameter $r\/$ ({\tt CW-WLS}$(r)$).

At the fractionary coordinates $(\xi, \eta)$ inside the $(K, L)$-th bin,  
\begin{equation}
   - 0.5~\leq~\xi~\le~0.5,~~- 0.5~\leq~\eta~\le~0.5,
\label{fracor}
\end{equation}
the {\tt CW-WLS}$(r)$ yields a smoothed value
\begin{equation}
  s_{K+\xi, L+\eta}~=~\sum_{K+k,L+l \in C_{KL}}~h_{K+k,L+l}~ G(\xi, \eta; k, l),
\label{smoothf}
\end{equation}
where the {\sl Green matrix\/} of the smoothing formula is given by
\begin{equation}
  G(\xi,\eta;k,l)~=~\sum_{m=0}^{M}~\nu^2_m~J_{0m}(k,l)~P_m(\xi,\eta).
\label{greenf}
\end{equation}
Here, $\nu_m$ denotes the norm of the $m$-th polynomial $P_m$ over its
definition area $C_{KL}$, while $J_{mm^\prime}(k,l)\/$, the overlap integral
of the basis polynomials $P_m(x,y)$
and $P_{m^\prime}(x,y)$ over the area of the $(k, l)$-th relative bin
inside the current smoothing window.
For a set of basis polynomials of at most 3-rd degree, the upper summation
index in Eq. (\ref{greenf}) is $M=10$.
\subsection{Optimal values of the window radial parameters}
\label{sec:optrp}

To achieve the {\sl best data fit\/} over an {\sl admissible class\/} of
functions, a consistent least squares fit procedure involves two kinds of
{\sl free parameters:\/} (i) the fit parameters which secure minimum $L_2$
norm departure of the data from a {\sl particular function\/} belonging to the
admissible class, and (ii) the free parameter which selects the {\sl best
function within the admissible class}. The difficult point is to fix the
last parameter of the procedure.

The {\tt CW-WLS} procedure for statistical noise smoothing described in
\cite{adam93a}
and summarized above solves the problem (i) (i.e., it defines the best
parameters of the {\sl constant polynomial degree\/} smoothing
filter~(\ref{smoothf}) under a given radial parameter, $r$, Eq.~(\ref{cklr})).
The solution of problem (ii), which consists in the definition of the
{\sl best $r$\/} value, that is of the most suitable smoothing window
$C_{KL}$ under an existing neighborhood of the reference bin $(K, L)$, will be
discussed below.

Within the usual {\sl polynomial fit}, which adjusts a
polynomial function of {\sl variable degree\/} to an input set consisting of
a {\sl fixed\/} number of elements, to get the {\sl best polynomial degree},
the method starts with a low polynomial degree which is then gradually
increased by unity until a stopping criterion is satisfied.

Of interest for the present investigation is the criterion proposed by Hamming
\cite{hamm73}.
Considering one-dimensional data, Hamming recommends the definition of the
best polynomial degree from the {\sl study of the distribution of the signs of
the residuals\/} of the smoothed data. When the polynomial degree is increased,
a threshold value is reached at which the signs of the residuals over the set
under study show {\sl stochastic distribution} (i.e., the residuals mainly
consist of statistical noise), while at higher polynomial degrees, the
{\sl prevalence of a given sign\/} over the set of residuals is obtained
(that is, a significant part of the useful signal is chopped by the
smoothing polynomial). The threshold defines the degree of the best fit
polynomial.

While the problem considered by Hamming is different from the present one, its
principle is very well suited to the definition of the radial parameter, $r$,
Eq.~(\ref{cklr}). Thus, instead of
varying the degree of the smoothing surface, we {\sl adjust its extension\/}
by variable $r$ values in a range $\{r_{min}, r_{max}\}$. We choose
$r_{min} = 2$, which is the lowest available radial parameter value for the
window (\ref{cklr}), while $r_{max} = 13$, a value which corresponds to the
chopping of the most part of the useful signal by the smoothing procedure.
A threshold value results, $r_{opt} \in \{r_{min}, r_{max}\}$, with similar
properties to those of the best polynomial degree within Hamming's procedure.

The residual of the smoothed value~(\ref{smoothf}) at the fractionary
$(\xi, \eta)$ coordinates inside the $(K+k, L+l)$-th bin belonging to the
window $C_{KL}$, Eq.~(\ref{cklr}) is, by definition,
\begin{equation}
  \delta_{K+k+\xi,L+l+\eta} = s_{K+k+\xi,L+l+\eta} - h_{K+k,L+l}.
\label{resid}
\end{equation}

Given a $C_{KL}$ window, signs of quantities (\ref{resid}) are computed over
two-dimensional submanifolds of it which obey to two requirements: the 
submanifolds have to be large enough such as to result in statistically
relevant information on the distribution of the signs of the residuals around
the reference bin, while low enough to secure a sufficiently fast algorithm.

The least possible manifold of interest is $V_1(K, L)$, the neighborhood of
the $(K, L)$-th bin which consists of the reference bin and its eight nearest
neighbors. If {\sl all\/} the nine signs entering $V_1(K, L)$ are
{\sl identical} with each other, then the distribution is assumed to be
non-stochastic.

If this rule fails, then we consider the larger neighborhood $V_2(K, L)$ of
the $(K,L)$-th bin, defined by the 21 bins of the $C_{KL}$ window~(\ref{cklr})
at the radial parameter value $r = r_{min} = 2$. If
{\sl two-thirds\/} at least of the signs of the residuals inside $V_2(K,L)$
are {\sl identical}, then the statistics is assumed to be non-stochastic.  

These two sign count criteria following from Hamming's procedure, however, do
not exhaust the manifold of non-stochastic two-dimensional distributions.
There is still the possibility of ordered sign distributions characterized
by nearly equal occurrence of positive and negative signs. To pick them out,
the analysis of {\sl directional sign distribution\/} of the residuals inside
$V_2(K,L)$ is performed along one or several of the following four directions:
the $p_x$ axis, the $p_y$ axis, the first bisectrix, and the second bisectrix.
In each case, we count the {\sl ordered triplets}, i.e., the sets of
{\sl identical three signs at neighboring bins\/} along rows, columns, or
diagonals respectively.  If the count yields a {\sl majority of ordered
triplets\/} with respect to the total number of possible ordered triplets in
the considered direction, then we decide that the sign distribution is
non-stochastic.

An {\sl accidental fulfillment\/} of one of the three abovementioned criteria
is possible, with the consequence that a spurious $r_{opt}$ cutoff is obtained.
To avoid such a premature end of the analysis, the requirement of
{\sl preservation\/} of the non-stochasticity of the sign distribution within
$V_1(K,L)$ or $V_2(K,L)$ under further increase of the radial parameter $r$
is imposed.

Within an ideal, perfect single crystal, the occurrence of jumps in the optimal
radial parameters closely follows the characteristic Fermi surface
jumps in the measured momentum distribution. Unfortunately, the real single
crystals show a great many number of imperfections of various kinds, which
significantly alter the obtained optimal window radial parameters. The crystal
imperfections blur the Fermi surface jump pattern and make it indistinguishable
from a stochastic momentum distribution. Thus, in real crystals, the
abovementioned definition of the optimal window parameters is to be
supplemented with a procedure able to restore the jumps at contiguous bins
which are characteristic to the Fermi surface.

Such a procedure is discussed in the next section.
\section{Elimination of Spurious Impurity Effects by Median Smoothing}
\label{sec:medsm}

The median smoothing is known to {\sl preserve the jumps}, if any, within
the distribution under consideration. Consistent median smoothing of physical
quantities associated to an $n$-axis-projected histogram, $H$, is obtained
provided the following two distinct problems are correctly solved.
The first concerns the {\sl parameters\/} of the two-dimensional smoothing
window: {\sl the extension of the neighborhood\/} of the $(K, L)$-th bin over
which the median smoothing is to be done and the {\sl weights\/} assigned to
the bins entering this neighborhood.
The second concerns the {\sl symmetry induced relationships\/} among the bins
of the histogram. Each of these problems is considered in detail below.

The {\sl extension of the neighborhood\/} entering the median smoothing
of the $(K, L)$-th bin can be chosen to be either a $V_1(K, L)$ or a
$V_2(K, L)$ neighborhood (defined in Sec.~\ref{sec:optrp} above).

Within a $V_1(K, L)$ neighborhood, it is natural to assume {\sl equal
weights\/} to all the bins entering it:
\begin{equation}
  \left(
    \matrix{
         1 & 1 & 1 \cr
         1 & 1 & 1 \cr
         1 & 1 & 1 \cr
	   }
  \right)
\label{m1}
\end{equation}

Within an $V_2(K, L)$ neighborhood, we can either assume {\sl equal
bin weights}
\begin{equation}
  \left(
    \matrix{
         0 & 1 & 1 & 1 & 0 \cr
         1 & 1 & 1 & 1 & 1 \cr
         1 & 1 & 1 & 1 & 1 \cr
         1 & 1 & 1 & 1 & 1 \cr
         0 & 1 & 1 & 1 & 0 \cr
	   }
  \right)
\label{m2}
\end{equation}
or we can assume {\sl different weights}. Here we consider the hat shape
\begin{equation}
  \left(
    \matrix{
         0 & 1 & 2 & 1 & 0 \cr
         1 & 3 & 5 & 3 & 1 \cr
         2 & 5 & 8 & 5 & 2 \cr
         1 & 3 & 5 & 3 & 1 \cr
         0 & 1 & 2 & 1 & 0 \cr
	   }
  \right)
\label{m4}
\end{equation}

In the case of data characterized by a high level of crystal imperfections,
the neighborhood $V_1(K, L)$ with equal weights (\ref{m1}) is too small to
result in effective cut of the fluctuations residing in the imperfections.
The neighborhood $V_2(K, L)$ with equal weights (\ref{m2}) is also
inappropriate since it overemphasizes the far staying bins, while the
weight of the reference bin is negligibly small. The most adequate seems to
be the choice $V_2(K, L)$ with unequal weights (\ref{m4}) which includes
sufficient enough neighbors while securing a fifty-fifty ratio between the
weights assigned to the reference bin and its nearest neighbours on one side
and the farther staying bins on the other side.

Since the information stored in different bins comes from a crystal, there
are strong symmetry induced correlations of the Fermi-surface-related
information stored in a {\tt 2D-ACAR} histogram. As a consequence, our
smoothing method has to be able to emphasize it. To implement this aspect
into the median smoothing, we make reference to the Neumann principle
(see, e.g.,
\cite{birss64}),
according to which, the symmetry originating in positron annihilations with
electrons within bands crossing the Fermi level, is a subduction to the
histogram plane of the symmetry point group of the reciprocal lattice of
the crystal.

For an $n$-axis-projected {\tt 2D-ACAR} histogram of interest here, the
distinct symmetry elements characterizing this electron fraction are: the
{\sl inversion symmetry center\/} $\Gamma $ (placed at the zero-momentum
projection of the distribution), a {\sl rotation axis\/} along {\bf n}, and
in-plane {\sl symmetry lines}.

The occurrence of the symmetry classifies the various bins entering an
$n$-axis-projected {\tt CF} histogram $H$ into {\sl stars of symmetry}.
In the case of an $YBa_2Cu_3O_{7-\delta}$ single crystal for instance, the
stars of the bins entering both the $c$-axis-projected and $a$-axis-projected
histograms contain one, two, or four elements, according to the fact that the
bin of interest contains the symmetry inversion center $\Gamma $ inside it, or
it lies along a symmetry axis, or it is a general bin.

In conclusion, there are three essential steps which secure the correct
approach to the median smoothing of an $n$-axis-projected {\tt 2D-ACAR}
histogram:
\begin{itemize}
 \item{
      Derivation of the {\tt CF} histogram $H$ from the {\tt LF} histogram
      {$\widetilde H$};
      }
 \item{
      Derivation of the window optimal radial parameters;
      }
 \item{
      Median smoothing over manifolds of data consisting of symmetry stars
      of bins, within neighborhoods $V_2$, with bin weights (\ref{m4}).
      }
\end{itemize}

Such a procedure finally yields window optimal radial parameters characteristic
to a structure showing the $n$-axis-projection of the Fermi surface of
the sample. As a consequence, if the accumulated statistics contains
characteristic Fermi surface jumps indeed, then the procedure should be
able to put them into evidence as {\sl characteristic jumps\/} in the final
window optimal radial parameters. That is, while a Fermi-surfaceless
structure is characterized by {\sl gradual\/} variations of the window
radial parameters (i.e., $\delta r$ jumps between neighboring bins in the
range \{-1, 0, 1\}), the occurrence of lines of Fermi surface within the
{\tt 2D} momentum projection should be evidenced by jumps
$\vert \delta r \vert \ge 2$ {\sl at contiguous bin positions}.

The next section illustrates this feature on a set of data open to
contradictory interpretation within the usual off-line processing methods.
\section{Results and Their Discussion}
\label{sec:appls}

The use of the present method is illustrated on a $c$-axis-projected
{\tt 2D-ACAR} histogram labelled "580", measured on an
$YBa_2Cu_3O_{7-\delta}$ single crystal, at room temperature, at the University
of Geneva, with an upgraded version of the setup described in
\cite{bisson82}
and previously analyzed with different methods of resolving the anisotropic
component of the {\tt 2D-ACAR} spectrum
\cite{adam93a,adam95,adam93b,hoffm93b,shukla95}.

The parameters which define the crystal frame $\Gamma p_xp_yp_z$
within the laboratory frame $Op_x^Dp_y^Dp_z^D$ are
\begin{equation}
   \{ \widetilde {\gamma }_x = \kappa _0 + \xi _0,
    ~~\widetilde {\gamma }_y = \lambda _0 + \eta _0,
    ~~\phi _0,~~\theta _0,~~\psi _0 \}. 
   \label{cfsol}
\end{equation}
Here ($\widetilde {\gamma }_x,\widetilde {\gamma }_y$) denote the in-plane
$\Gamma $ coordinates, while $\phi _0,~\theta _0,~\psi _0$ are the Euler
angles which define the rotations from the $Op_x^Dp_y^Dp_z^D$ frame axes
to the $\Gamma p_xp_yp_z$ frame axes.

The rough approximations $(\kappa _0, \lambda _0)$, to the coordinates of
$\Gamma $ have been found to be $(-1.0, 5.0)$, while the fractionary
coordinates were found to be $\xi _0 = -0.1146\times 10^{-1}$ and
$\eta _0 = 0.2413$
(in $a_D$ units). The value of the Euler angle $\theta _0$ was
found to be negligibly small ($\theta _0 = -0.316\times 10^{-3}$ radians),
such that the experimental histogram can be accepted for a
$c$-axis-projected histogram indeed. Then the only relevant Euler angle
is the sum of the angles $\phi _0$ and $\psi _0$,
$\alpha _0 = -0.12297\times 10^{-2}$ radians.

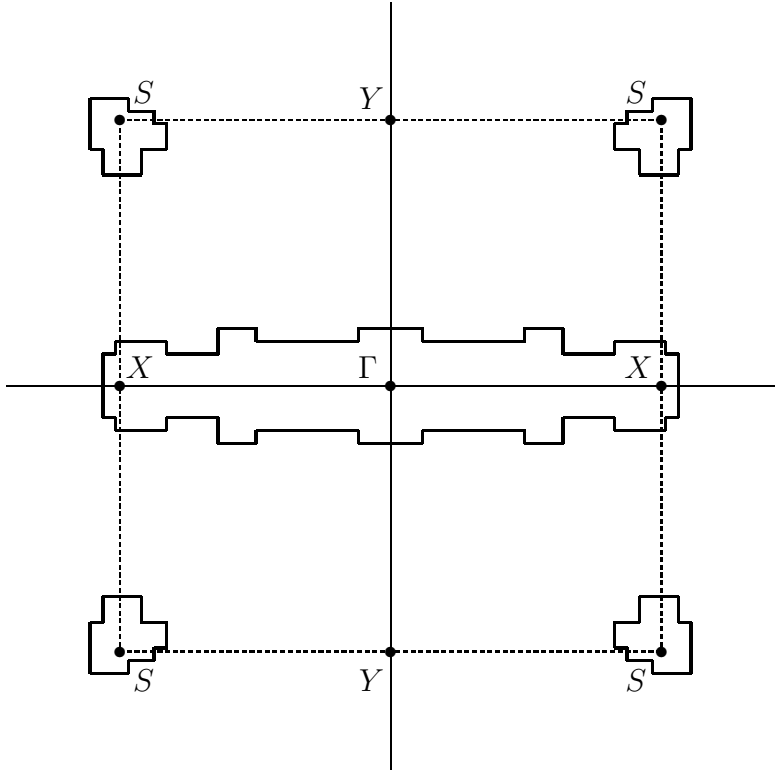
\begin{figure}[h]
\begin{center}
\setlength{\unitlength}{1.7mm}
\begin{picture}(60,60)
         \thinlines
         \put(0,30){\line(1,0){60}}
         \put(30,0){\line(0,1){60}}
         \put(30,30){\circle*{0.75}}
         \multiput(8.81,30)(42.38,0){2}{\circle*{0.75}}
         \multiput(30,9.19)(0,41.62){2}{\circle*{0.75}}
         \multiput(8.81,9.19)(42.38,0){2}{\circle*{0.75}}
         \multiput(8.81,50.81)(42.38,0){2}{\circle*{0.75}}
         \put(8.81,9.19){\dashbox{0.3}(42.38,41.62)}
          \thicklines
         \multiput(6.5,7.5)(47,0){2}{\line(0,1){4}}
         \multiput(7.5,11.5)(45,0){2}{\line(0,1){2}}
         \multiput(9.5,7.5)(41,0){2}{\line(0,1){1}}
         \multiput(10.5,11.5)(39,0){2}{\line(0,1){2}}
         \multiput(11.5,8.5)(37,0){2}{\line(0,1){1}}
         \multiput(12.5,9.5)(35,0){2}{\line(0,1){2}}
         \multiput(6.5,48.5)(47,0){2}{\line(0,1){4}}
         \multiput(7.5,46.5)(45,0){2}{\line(0,1){2}}
         \multiput(9.5,51.5)(41,0){2}{\line(0,1){1}}
         \multiput(10.5,46.5)(39,0){2}{\line(0,1){2}}
         \multiput(11.5,50.5)(37,0){2}{\line(0,1){1}}
         \multiput(12.5,48.5)(35,0){2}{\line(0,1){2}}
         \multiput(6.5,7.5)(44,0){2}{\line(1,0){3}}
         \multiput(6.5,11.5)(46,0){2}{\line(1,0){1}}
         \multiput(7.5,13.5)(42,0){2}{\line(1,0){3}}
         \multiput(9.5,8.5)(39,0){2}{\line(1,0){2}}
         \multiput(10.5,11.5)(37,0){2}{\line(1,0){2}}
         \multiput(11.5,9.5)(36,0){2}{\line(1,0){1}}
         \multiput(6.5,52.5)(44,0){2}{\line(1,0){3}}
         \multiput(6.5,48.5)(46,0){2}{\line(1,0){1}}
         \multiput(7.5,46.5)(42,0){2}{\line(1,0){3}}
         \multiput(9.5,51.5)(39,0){2}{\line(1,0){2}}
         \multiput(10.5,48.5)(37,0){2}{\line(1,0){2}}
         \multiput(11.5,50.5)(36,0){2}{\line(1,0){1}}
         \put(29,31.5){\makebox(0,0)[r]{$\Gamma$}}
         \multiput(11.5,31.5)(39,0){2}{\makebox(0,0)[r]{$X$}}
         \multiput(29.5,7.0)(0,45.5){2}{\makebox(0,0)[r]{$Y$}}
         \multiput(11.5,7.0)(38.5,0){2}{\makebox(0,0)[r]{$S$}}
         \multiput(11.5,53)(38.5,0){2}{\makebox(0,0)[r]{$S$}}
         \multiput(7.5,27.5)(45,0){2}{\line(0,1){5}}
         \multiput(8.5,26.5)(43,0){2}{\line(0,1){1}}
         \multiput(8.5,32.5)(43,0){2}{\line(0,1){1}}
         \multiput(12.5,26.5)(35,0){2}{\line(0,1){1}}
         \multiput(12.5,32.5)(35,0){2}{\line(0,1){1}}
         \multiput(16.5,25.5)(27,0){2}{\line(0,1){2}}
         \multiput(16.5,32.5)(27,0){2}{\line(0,1){2}}
         \multiput(19.5,25.5)(21,0){2}{\line(0,1){1}}
         \multiput(19.5,33.5)(21,0){2}{\line(0,1){1}}
         \multiput(27.5,25.5)(5,0){2}{\line(0,1){1}}
         \multiput(27.5,33.5)(5,0){2}{\line(0,1){1}}
         \multiput(7.5,27.5)(44,0){2}{\line(1,0){1}}
         \multiput(7.5,32.5)(44,0){2}{\line(1,0){1}}
         \multiput(8.5,26.5)(39,0){2}{\line(1,0){4}}
         \multiput(8.5,33.5)(39,0){2}{\line(1,0){4}}
         \multiput(12.5,27.5)(31,0){2}{\line(1,0){4}}
         \multiput(12.5,32.5)(31,0){2}{\line(1,0){4}}
         \multiput(16.5,25.5)(24,0){2}{\line(1,0){3}}
         \multiput(16.5,34.5)(24,0){2}{\line(1,0){3}}
         \multiput(19.5,26.5)(13,0){2}{\line(1,0){8}}
         \multiput(19.5,33.5)(13,0){2}{\line(1,0){8}}
         \multiput(27.5,25.5)(0,9){2}{\line(1,0){5}}
\end{picture}
\label{fig:pillb}
\caption{
        Occurrence of jumps of the window optimal radial parameters of the
        {\tt CW-WLS} method for statistical noise smoothing resolves two
        Fermi surface sheets: the {\sl ridge}, going from $\Gamma $ to $X$
        across the whole first Brillouin zone of the sample and the
        {\sl pillboxes}, around the corners $S$ of the first Brillouin zone.
        }
\end{center}
\end{figure}

The final result of the procedure described in the previous sections is
shown in figure~\ref{fig:pillb}. Both the electron ridge crossing the first
Brillouin zone from $\Gamma$ to $X$ and the pillbox around the $S$ corners
are resolved. It is to be stressed that the ridge-jumps have been of high
amplitude ($\delta r$ inbetween three to six units large), while the
pillbox-jumps have always been characterized by the minimal $\delta r = 2$ 
value, telling us that the collected ridge jump signal is strong, while the
pillbox jump signal is weak.

Whereas the occurrence of the Fermi surface ridge in the analyzed "580"
histogram was confirmed by different signal processing techniques in all
the five references mentioned above,
\cite{adam93a,adam95,adam93b,hoffm93b,shukla95}.
the only references which analyze the pillbox are
\cite{shukla95} 
and
\cite{adam95}.

The methods of analysis used in
\cite{shukla95})
did not confirm the occurrence of the pillbox in $YBa_2Cu_3O_{7-\delta}$
at room temperature. The measurements of the positron mean free path in
this compound at different temperatures suggested that, at room temperature,
there is a shallow positron trapping mechanism which is responsible of
the absence of the pillbox.

Our previous analysis of the same spectrum
\cite{adam95}
has found that the resolved pillbox exists, but it is small (the
depth of the Lock-Crisp-West ({\tt LCW}) folded
\cite{lcw73}
signal at the $S$ point was found to be only four times larger than the
experimental errors).

The present investigation, which is not symmetric component subtraction
sensitive, adds considerable weight to the conclusions of reference
\cite{adam95},
without making use of the {\tt LCW} folding to achieve signal enhancement.
It lets us infer that, while weakening considerably the pillbox signature at
room temperature indeed, the positron shallow trapping does not rule it out
altogether. It is to be noted, however, that, although describing a closed
contour jump in the measured positron-electron momentum distribution, the shape
of the resolved pillbox around the $S$ point is irregular, a feature which
tells us about is small amplitude, which can be easily distorted by the
occurrence of non-homogeneously distributed crystal defects.
\section{Acknowledgments}
We are very much indebted to Professor Martin Peter from the University of
Geneva, Switzerland, who introduced us to the field of positron spectroscopy.
Discussion with him and with Professor Alfred A. Manuel from Geneva University
at various stages of the investigation, as well as provision of the
experimental data, are gratefully acknowledged.

Part of this work was done within the framework of the Associateship Scheme of
the Abdus Salam International Centre for Theoretical Physics, Trieste, Italy.
The first author would like to thank Professor M. Virasoro and the Abdus Salam
ICTP for hospitality and for granting him generous access to the computing
facilities of the Centre.

The work done in Romania was financed by the Ministry of Research
and Technology of Romania.
%
%


\begin{thebibliography}{99}
\bibitem{berko83}{S. Berko, in: Positron Solid State Physics, Proc. of
the Int.  School \lq\lq E. Fermi\rq\rq, Course 83, W. Brandt and A. Dupasquier
eds., (North Holland, New York, 1983) pp. 64--145. Reprinted in
Positron Studies of Solids, Surfaces and Atoms, A.P. Mills,Jr.,
W.S. Crane and K.F. Canter eds., (World Scientific, Singapore, 1986)
pp. 246--327.}
\bibitem{berko89}{
S.Berko, in: Momentum Distributions, R.N. Silver and P.E. Sokol eds.,
(Plenum Press, New York, 1989) p. 273.}
\bibitem{peter89}{
M. Peter, IBM J. Res. Develop. 33/3 (1989), 333.}
\bibitem{chan92}{
L.P. Chan, K.G. Lynn and D.R. Harshman, Mod. Phys. Lett. B, 6 (1992), 617,
and references therein.}
\bibitem{pankal94}{
R. Pankaluoto, A. Bansil, L.C. Smedskjaer and P.E. Mijnarends,
Phys. Rev. B, 50 (1994), 6408.}
\bibitem{adam93a}{
Gh. Adam, S. Adam, B. Barbiellini, L. Hoffmann, A.A. Manuel and M. Peter,
Nucl. Instr. and Meth. A, 337 (1993), 188.}
\bibitem{hoffm93a}{
L. Hoffmann, A. Shukla, M. Peter, B. Barbiellini and A.A. Manuel,
Nucl. Instr. and Meth. A, 335 (1993), 276.}
\bibitem{west95}{
R.N. West, in: Positron Spectroscopy of Solids, A. Dupasquier and
A.P. Mills, Jr. eds., (North Holland, New York, 1995).}
\bibitem{hoffm91}{
L. Hoffmann, W. Sadowski, A. Shukla, Gh. Adam, B. Barbiellini, and
M. Peter, J. Phys. Chem. Sol. 52 (1991), 1551.}
\bibitem{smed92}{
L.C. Smedskjaer, A. Bansil, U. Welp, Y. Fang and K.G. Bailey,
Physica C, 192 (1992), 259.}
\bibitem{bradl72}{          
C.J. Bradley and A.P. Cracknell, The Mathematical Theory of
Symmetry in Solids: Representation Theory for Point Groups and Space
Groups (Clarendon Press, Oxford, 1972).}
\bibitem{adam95}{
Gh. Adam and S. Adam, International J. Modern Phys. B, 9 (1995), 3667.}
\bibitem{adam93c}{
Gh. Adam and S. Adam, Romanian J. Phys., 38 (1993), 681.} 
\bibitem{hamm73}{
R. W. Hamming, Numerical Methods for Scientists and Engineers, 2-nd ed.
(Mc Graw-Hill, New York, 1973), Chaps.~25--27.} 
\bibitem{birss64}{
R.R. Birss, Symmetry and Magnetism (North-Holland, Amsterdam, 1964).}
\bibitem{bisson82}{
P.E. Bisson, P. Descouts, A. Dupanloup, A.A. Manuel, E. Perreard,
M. Peter and R. Sachot, Helv. Phys. Acta 55 (1982), 100.}
\bibitem{adam93b}{
Gh. Adam, S. Adam, B. Barbiellini, L. Hoffmann, A.A. Manuel, M. Peter
and S. Massidda, Solid State Commun. 88 (1993), 739.}
\bibitem{hoffm93b}{
L. Hoffmann, A.A. Manuel, M. Peter, E. Walker, M. Gauthier, A. Shukla,
B. Barbiellini, S. Massidda, Gh. Adam, S. Adam, W.N. Hardy, Ruixing Liang,
Phys. Rev. Lett. 71 (1993), 4047.}
\bibitem{shukla95}{
A. Shukla, L. Hoffmann, A.A. Manuel, E. Walker, B. Barbiellini, and
M. Peter, Phys. Rev. B 51 (1995), 6028.}
\bibitem{lcw73}{            
D.G. Lock, V.H.C. Crisp and R.N. West, J. Phys. F: Metal Phys. 3 (1973), 561.} 
\end{thebibliography}
\end{document}